\documentclass[pra,aps,longbibliography,twocolumn,superscriptaddress,floatfix,footnoteinbib]{revtex4-1}
\usepackage{amsmath,amssymb,graphicx,units}
\usepackage[usenames,dvipsnames]{color}
\usepackage{lipsum}
\usepackage{setspace}
\usepackage{ulem}
\usepackage[bookmarks=false,linkcolor=blue,urlcolor=blue,colorlinks,citecolor=blue]{hyperref}

\usepackage{mathrsfs}

\begin{document}
\title{Variational generalized rotating-wave approximation in the two-qubit quantum Rabi model}

\author{Bin-Bin Mao}
\affiliation{School of Physical Science and Technology $\&$ Key Laboratory for Magnetism and Magnetic Materials of the MoE, Lanzhou University, Lanzhou 730000, China}

\author{Liangsheng Li}
\affiliation{Science and Technology on Electromagnetic Scattering Laboratory, Beijing 100854, China}

\author{Yimin Wang}
\affiliation{Communications Engineering College, Army Engineering University, Nanjing 210007, China}
\affiliation{Shanghai Key Laboratory of High Temperature Superconductors, Shanghai 200444, China}
\author{Wen-Long You}
\affiliation{School of Physical Science and Technology, Soochow University, Suzhou, Jiangsu 215006, China}

\author{Wei Wu}
\affiliation{Beijing Computational Science Research Center, Beijing 100094, China}

\author{Maoxin Liu}
\email{liumaoxin@bupt.edu.cn}
\affiliation{
State Key Laboratory of Information Photonics and Optical
Communications $\&$ School of Science, Beijing University of Posts and
Telecommunications, Beijing 100876, China}
\affiliation{Beijing Computational Science Research Center, Beijing 100094, China}

\author{Hong-Gang Luo}
\email{luohg@lzu.edu.cn}
\affiliation{School of Physical Science and Technology $\&$ Key Laboratory for Magnetism and Magnetic Materials of the MoE, Lanzhou University, Lanzhou 730000, China}
\affiliation{Beijing Computational Science Research Center, Beijing 100094, China}
	
	
\begin{abstract}
We present an analytical method for the two-qubit quantum Rabi model. While still operating in the frame of the generalized rotating-wave approximation (GRWA), our method further embraces the idea of introducing variational parameters. The optimal value of the variational parameter is determined by minimizing the energy function of the ground state. Comparing with numerical exact results, we show that our method evidently improves the accuracy of the conventional GRWA in calculating fundamental physical quantities, such as energy spectra, mean photon number, and dynamics. Interestingly, the accuracy of our method allows us to reproduce the asymptotic behavior of mean photon number in large frequency ratio for the ground state  and investigate the quasi-periodical structure of the time evolution, which are incapable of being predicted by the GRWA. The applicable parameter ranges cover the ultrastrong coupling regime, which will be helpful to recent experiments.
\end{abstract}
	
\maketitle
	
\section{Introduction}\label{intro}
The quantum Rabi model~\cite{rabi1936,rabi1937} describes a two-level system linearly interacting with a single mode bosonic field. It plays a fundamental role in many areas of physics, such as quantum optics~\cite{Herbert2006}, quantum information~\cite{Raimond2001}, condensed matter physics~\cite{Holstein1959}. The history of the quantum Rabi model can be traced back to more than 80 years ago, when the original version of semi-classical Rabi model was introduced. Recently, the model has attracted much attention due to the fact that the so-called ultrastrong coupling~\cite{Bourassa2009,Niemczyk2010,Peropadre2010,Ridolfo2012,Todorov2014,Baust2016,Forn2016}, and even deep strong coupling~\cite{Casanova2010,Liberato2014,FornDiaz2017,Yoshihara2017,Chen2017} regime have been experimentally achieved. When the qubit-oscillator coupling strength is strong enough, the counter-rotating terms in the model can no longer be ignored.
For example, the experimental observation of the Bloch-Siegert shift~\cite{Forn2010} emphasizes the importance of the counter-rotating term.
To this end, a series of fascinating phenomena have been explored in the model without rotating-wave approximation (RWA)~\cite{Jaynes1963,Shore1993}, e. g., generation of photons~\cite{Werlang2008}, entanglement from zero excitation initial state~\cite{Burnett2008}, bifurcation in the phase space~\cite{Simon1991}, a fine structure in the optical Stern Gerlach effect~\cite{Lembessis2008}.
In particular, it has recently been noted that in the frequency ratio limit, the model undergoes a superradiant phase transition~\cite{Hwang2015,Liu2017}. These experimental and theoretical progresses fascinate one to further explore the quantum Rabi model and the related issues.

The theoretical starting point to study the quantum Rabi model is to solve the eigenvalues of the model Hamiltonian. Despite the simple form of the model, it was only until the year 2011 D. Braak has obtained its integrability~\cite{Braak2011}. Comparing to the great achievement of the exact solution in mathematical aspect, extracting the physical information of the model by this solution is still a non-trivial task. Thus, people have been still trying to develop a variety of analytical approximations such as adiabatic approximation~\cite{Ashhab2010,Agarwal2012,Irish2005}, RWA~\cite{Jaynes1963,Shore1993,Irish2005}, generalized rotating-wave approximation (GRWA)~\cite{Irish2007}, extended coherent-state
method~\cite{Mao2015}, continued fraction~\cite{Swain1973}, perturbation method~\cite{Yu2012}.
These methods are widely and fruitfully used in calculating the single-qubit quantum Rabi model. In parallel, some efforts have been devoted to its multi-qubit counterpart.
In fact, the multi-qubit version of the quantum Rabi model is of great value in both theoretical studies and practical applications.
The $N$-qubit version of quantum Rabi model is well known as the Dicke model, in which a famous superradiant phase transition occurs~\cite{Dicke1954}. In application sense, one needs multi-qubit setups to do quantum computing. For example, implementing quantum gate operations requires at least two qubits~\cite{Zhou2006,Huang2008}. There are also many interesting issues involving in multi-qubit scenario, i.e., genuine multipartite entanglement~\cite{Ollivier2001,Zurek2003,Roscilde2004,Horodecki2009,Jungnitsch2011,Giampaolo2013}, quantum simulation of dynamical maps~\cite{Schindler2013}, and holonomic quantum computation~\cite{Wang2016}. Therefore, in this work, to explore the multi-qubit effects, we focus on the two-qubit quantum Rabi model, which has been relatively less studied.

An adiabatic approximation has been proposed for the two-qubit quantum Rabi model~\cite{Agarwal2012}.
It plays well when the qubit frequency is much smaller than the oscillator frequency, and has been improved by the GRWA when the two frequencies are comparable~\cite{Zhang2015}. However, recently, another frequency ratio regime, i. e., the qubit frequency is larger than the oscillator frequency, attracts much interest because the quantum phase transition occurs in this frequency ratio limit~\cite{Hwang2015,Liu2017}. In such a case, we find that the GRWA still has a space to be improved. For example, the physical observable in ultrastrong coupling regime predicted by the GRWA is not accurate enough, and the dynamic process calculated by the GRWA misses the quasi-periodical structures. The variational method in Ref.~\cite{Lee2012}
performs well in improving the GRWA. However, it is only limited to the ground state. Therefore, in this work, we try to improve the GRWA in the variational scheme and consider both ground state and low excited states. The main idea is to introduce a variational parameter which is optimized by
minimizing the energy function for the ground state. To show the advantage of our method, all of calculated results by the GRWA are compared with the exact diagonalization results as a benchmark.

The paper is organized as follows. In Section~\ref{sec_2}, the two-qubit quantum Rabi model is  introduced.
We then show the detail of our analytical method to obtain the eigenvalues and eigenstates of the two-qubit quantum Rabi model. In Section~\ref{sec_4}, we calculate several physical quantities and compare them obtained by GRWA with those obtained by exact diagonalization. Section~\ref{sec_5} gives a brief summary.
	
\section{Model and analytical solution}\label{sec_2}
The Hamiltonian of the two-qubit quantum Rabi model reads as
\begin{equation}\label{eq_hamiltonian}
\hat{H}=\omega \hat{a}^{\dagger}\hat{a}+\Omega \hat{J}_{x}+g\hat{J}_{z}\left(\hat{a}^{\dagger}+\hat{a}\right),
\end{equation}
where $\hat{a}^{\dagger}\left(\hat{a}\right)$ is the creation (annihilation) operator of the harmonic oscillator with frequency $\omega$, $\Omega$ is the atomic transition frequency and $g$ denotes the coupling strength between qubits and oscillator. The angular momentum operators can be assembled by the Pauli matrix of two identical qubits as $\hat{J}_{x}=\frac{1}{2}\left(\hat{\sigma}_{x}^{1}+\hat{\sigma}_{x}^{2}\right)$, $\hat{J}_{y}=\frac{1}{2}\left(\hat{\sigma}_{y}^{1}+\hat{\sigma}_{y}^{2}\right)$, $\hat{J}_{z}=\frac{1}{2}\left(\hat{\sigma}_{z}^{1}+\hat{\sigma}_{z}^{2}\right)$
and 1,2 represent each of two qubits. 
Note that the total spin operator $\hat{J}^2$
commutates with the Hamiltonian in Eq. \eqref{eq_hamiltonian}, and the spin singlet state($J=0$) is decoupled from the bosonic mode. Thus we only consider the spin triplet states($J=1$) in this work. 
In the following calculations, we take $\omega=1$ as an energy scale.
Despite the fact of still working in the frame of the GRWA, the key idea of  our method is to choose an optimal variational parameter,　which minimizes the energy function of the ground state. 
In the following we will exhibit our variational scenario step by step. During such procedure, the adiabatic approximation and the GRWA will be explicitly recovered as leading-order approximations.

To start, we make a unitary transformation onto the Hamiltonian, i.e., $\hat{\tilde{H}} = \hat{U}\hat{H}\hat{U}^{\dagger}$.
Similar to the GRWA~\cite{Irish2007,Zhang2015}
, we choose $\hat{U}=e^{\lambda\hat{J}_{z}\left(\hat{a}^{\dagger}-\hat{a}\right)}$ 
. The difference is that $\lambda$ is undetermined here rather than that is fixed to be $g/\omega$ in the GRWA. More explicitly, there is
\begin{equation}\label{uham}
\begin{split}
\hat{\tilde{H}}\! =& \hat{U}\hat{H}\hat{U}^{\dagger} \\
 =& \omega \hat{a}^{\dagger}\hat{a}+\left(\lambda^{2}\omega-2g\lambda\right)\hat{J}_{z}^{2}+ \left(g-\lambda\omega\right)\hat{J}_{z}\left(\hat{a}^{\dagger}+\hat{a}\right)\\
 &+ \Omega \hat{J}_{x}\mathrm{cosh}\left[\lambda(\hat{a}^{\dagger}-\hat{a})\right]+ {\rm i}\Omega J_{y}\mathrm{sinh}\left[\lambda(\hat{a}^{\dagger}-\hat{a})\right],
\end{split}
\end{equation}
where the hyperbolic sine and cosine terms can be further expanded as
\begin{equation}
\begin{split}
&\mathrm{sinh}\left[\lambda\left(\hat{a}^{\dagger}-\hat{a}\right)\right]\\
=&\sum_{k=0}^{\infty}\left[\left(\hat{a}^{\dagger}\right)^{2k+1}F_{2k+1}\left(\hat{a}^{\dagger}\hat{a}\right)+F_{2k+1}\left(\hat{a}^{\dagger}\hat{a}\right)\hat{a}^{2k+1}\right],
\end{split}
\end{equation}
and
\begin{equation}
\begin{split}
&\mathrm{cosh}\left[\lambda\left(\hat{a}^{\dagger}-\hat{a}\right)\right]\\
=&F_{0}\!\left(\hat{a}^{\dagger}\hat{a}\right)\!+\!\sum_{k=1}^{\infty}\bigg[\!\left(\hat{a}^{\dagger}\right)^{2k}F_{2k}\left(\hat{a}^{\dagger}\hat{a}\right)\!+\!F_{2k}\left(\hat{a}^{\dagger}\hat{a}\right)\hat{a}^{2k}\bigg],
\end{split}
\end{equation}
respectively.
The function $F_{m}$ is defined as
\begin{equation}
\begin{split}
F_{m}\left({n}\right)=e^{-\lambda^{2}/2}\lambda^{m}\frac{n!}{\left(n+m\right)!}L_{n}^{m}\left(\lambda^{2}\right),
\end{split}
\end{equation}
where $m$ and $n$ are integers, $L_{n}^{m}\left(x\right)=\sum_{i=0}^{n}\left(-x\right)^{i}\frac{\left(n+m\right)!}{\left(m+i\right)!\left(n-i\right)!i!}$ is the associated Laguerre polynomial.
Although Hamiltonian in Eq.~\eqref{uham} is still hard to solve, one can further employ some approximations.

A zero-order approximation of Eq.~\eqref{uham} is made in the so-called adiabatic approximation, where spin and oscillator are decoupled.
That is,

\begin{equation}
\hat{\tilde{H}}_{\mathrm{0}}  =   \omega \hat{a}^{\dagger}\hat{a}+\Omega \hat{J}_{x}F_{0}\left(\hat{a}^{\dagger}\hat{a}\right)
+\left(\lambda^{2}\omega-2g\lambda\right)\hat{J}_{z}^{2}.
\end{equation}
Taking the direct product basis  $\left|j_{x}\right\rangle \otimes \left | n\right\rangle$, where $\left|j_{x}\right\rangle$ is the eigenstate of $\hat{J}_{x}$ and $\left | n\right\rangle$ is the Fock state of the oscillator, the Hamiltonian can be written in each isolate $n$-dependent subspace in terms of a $3\times 3$ matrix as
\begin{equation}\label{aamtx}
	\hat{\tilde{H}}_{\mathrm{0}}\left(n,\lambda\right)  =  \left(\begin{array}{ccc}
		\xi_{n}^{-} & 0 &  \epsilon_{\lambda}\\
		0 & \xi_{n}^{0} & 0\\
		\epsilon_{\lambda} & 0 & \xi_{n}^{+}
	\end{array}\right),
\end{equation}
where $\xi_{n}^{\pm}=\epsilon_{n}+\epsilon_{\lambda}\pm f_{n}^{0}$, $\xi_{n}^{0}=\epsilon_{n}+2\epsilon_{\lambda}$, $\epsilon_{n}=\omega n$, $\epsilon_{\lambda}=\left(\lambda^{2}\omega-2g\lambda\right)/2$, $f_{n}^{0}=\Omega F_{0}(n)$.
The Hamiltonian block in Eq.~\eqref{aamtx} can be analytically solved (see Appendix \ref{appendix_aa}).

If $\lambda = g/\omega$ is assumed, the solution of Eq.~\eqref{aamtx} will retrieve the result in Ref.~\cite{Agarwal2012}. We note that our general adiabatic approximation can perform better than Ref.~\cite{Agarwal2012}, since $\lambda$ can be further optimized by minimizing the ground-state energy (see Appendix \ref{appendix_aa}). Before discussing how to choose an optimal value of $\lambda$, we will then consider higher-order terms in Hamiltonian for a more accurate approximation.

In the spirit of the GRWA, we consider the Hamiltonian up to single-excitation terms, so that we write
\begin{equation}
\hat{\tilde{H}}_{\mathrm{1}}  = \hat{\tilde{H}}_{\mathrm{0}}+\hat{\tilde{H}}_{\mathrm{GRW}}+\hat{\tilde{H}}_{\mathrm{GCRW}},
\end{equation}
where
\begin{equation}
\begin{split}
\hat{\tilde{H}}_{\mathrm{GRW}} =&\frac{1}{2}\left(g-\lambda\omega\right)\left(\hat{J}_{+}\hat{a}+\hat{J}_{-}\hat{a}^{\dagger}\right)\\
&+\frac{1}{2}\Omega \left[\hat{J}_{+}F_{1}\left(\hat{a}^{\dagger}\hat{a}\right)\hat{a}+\hat{J}_{-}\hat{a}^{\dagger}F_{1}\left(\hat{a}^{\dagger}\hat{a}\right)\right],
\end{split}
\end{equation}
is the generalized rotating-wave term, which conserves the total excitation, and
\begin{equation}
\begin{split}
\hat{\tilde{H}}_{\mathrm{GCRW}}  =&\frac{1}{2}\left(g-\lambda\omega\right)\left(\hat{J}_{-}\hat{a}+\hat{J}_{+}\hat{a}^{\dagger}\right)\\
&-\frac{1}{2}\Omega \left[\hat{J}_{-}F_{1}\left(\hat{a}^{\dagger}\hat{a}\right)\hat{a}+\hat{J}_{+}\hat{a}^{\dagger}F_{1}\left(\hat{a}^{\dagger}\hat{a}\right)\right],
\end{split}
\end{equation}
is the generalized counter-rotating-wave term, which is the counterpart of $\hat{\tilde{H}}_{\mathrm{GRW}}$.
Here $\hat{J}_{\pm}=\frac{1}{2}(\hat{J}_{z}\mp i\hat{J}_{y})$ and $\hat{a}$ ($\hat{a}^{\dagger}$) are single-excitation operators for spin and oscillator, respectively.

To let $\hat{\tilde{H}}_{\mathrm{1}}$ be solvable, $\hat{J}_{-}^{2}$ and $\hat{J}_{+}^{2}$ terms in $\hat{\tilde{H}}_{\mathrm{0}}$ should be neglected, since they correspond to the remote off-diagonal entries in the spin representation $\left|j_{x}\right\rangle$. We also eliminate the generalized counter-rotating term $\hat{\tilde{H}}_{\mathrm{GCRW}}$, such that the total excitation of the Hamiltonian is conserved. Then, we obtain a solvable one named GRWA Hamiltonian:
\begin{widetext}
\begin{equation}\label{grwah}
\hat{\tilde{H}}_{\mathrm{GRWA}}  = \omega \hat{a}^{\dagger}\hat{a}+\Omega \hat{J}_{x}F_{0}\left(\hat{a}^{\dagger}\hat{a}\right)+\frac{1}{4}\left(\lambda^{2}\omega-2g\lambda\right)\left(\hat{J}_{+}\hat{J}_{-}+\hat{J}_{-}\hat{J}_{+}\right)\\
+\frac{1}{2}\left[\hat{J}_{+}\left(g-\lambda\omega+\Omega F_{1}\left(\hat{a}^{\dagger}\hat{a}\right)\right)\hat{a}+h.c.\right].
\end{equation}
\end{widetext}
If one chooses $\lambda=g/\omega$, the Hamiltonian in Eq.~\eqref{grwah} recovers the GRWA in Ref.~\cite{Zhang2015}.

The Hamiltonian $\hat{\tilde{H}}_{\mathrm{GRWA}}$ is a block-diagonal matrix in the bases subspace \{$\left|1_{x}\right\rangle \otimes \left| n-1\right\rangle$, $\left|0_{x}\right\rangle \otimes \left|n\right\rangle$, $\left|-1_{x}\right\rangle \otimes \left|n+1\right\rangle$\}.
The $n$-th block takes the form of
\begin{widetext}
\begin{equation}\label{eq:H_matrix}
\hat{\tilde{H}}^{\prime}_{n}(\lambda)=\left(\begin{array}{ccc}
\omega\left(n-1\right)+f_{n-1}^{0}+\epsilon_{\lambda} & \sqrt{\frac{n}{2}}\left(f_{n-1}^{1}+\lambda^{\prime}\right) & 0\\
\sqrt{\frac{n}{2}}\left(f_{n-1}^{1}+\lambda^{\prime}\right) & \omega n+2\epsilon_{\lambda} & \sqrt{\frac{n+1}{2}}\left(f_{n}^{1}+\lambda^{\prime}\right)\\
0 & \sqrt{\frac{n+1}{2}}\left(f_{n}^{1}+\lambda^{\prime}\right) & \omega\left(n+1\right)-f_{n+1}^{0}+\epsilon_{\lambda}
\end{array}\right),
\end{equation}
\end{widetext}
where $\epsilon_{\lambda}=\left(\lambda^{2}\omega-2g\lambda\right)/2$, $\lambda^{\prime}=g-\lambda\omega$, $f_{n}^{1}=\Omega F_{1}(n)$.
The eigenvalues $E_{n}^{j}\left(j=\{1,2,3\}\right)$ and the corresponding eigenvectors $\overrightarrow{\tilde{\psi}}_{n}^{j}$ can be obtained as
\begin{equation}
\hat{\tilde{H}}_{n}^{\prime}(\lambda)\overrightarrow{\tilde{\psi}}_{n}^{j}=E_{n}^{j}\overrightarrow{\tilde{\psi}}_{n}^{j}.
\end{equation}
Thus the wavefunctions can be expressed as
\begin{equation}
\left|\tilde{\phi}_{n}^{j}\right\rangle=\left(\overrightarrow{\tilde{\psi}}_{n}^{j}\right)^{T}\left(\begin{array}{c}
\left|1_{x},n-1\right\rangle \\
\left|0_{x},n\right\rangle \\
\left|-1_{x},n+1\right\rangle
\end{array}\right),
\end{equation}
and take the form
\begin{equation}
\begin{split}
\left|\tilde{\phi}_{n}^{j}\right\rangle
= & c_{1,n}^{j}\left|1_{x}\right\rangle \otimes \left|n-1\right\rangle+c_{0,n}^{j}\left|0_{x}\right\rangle \otimes \left|n\right\rangle\\
&+c_{-1,n}^{j}\left|-1_{x}\right\rangle \otimes \left|n+1\right\rangle,
\end{split}
\end{equation}
where the coefficients $\{c_{m_{s},n}^{j}\}$ are given in Appendix \ref{appendix_energy} in detail.

There is a special case for $n = 0$. In the bases $\{\left|0_{x}\right\rangle \otimes \left|0\right\rangle, \left|-1_{x}\right\rangle \otimes \left|1\right\rangle\}$, we have
\begin{equation}
\begin{split}
\hat{\tilde{H}}_{0}^{\prime} (\lambda)
=&\left(\begin{array}{cc}
\varepsilon_{0}^{0}& R_{0,1}  \\
R_{0,1}   & \varepsilon_{1}^{-}
\end{array}\right),
\end{split}
\end{equation}
where $\varepsilon_{0}^{0}=2\epsilon_{\lambda}$, $R_{0,1}=\sqrt{\frac{1}{2}}\left(f_{0}^{1}+\lambda^{\prime}\right)$, $\varepsilon_{1}^{-}=\omega-f_{1}^{0}+\epsilon_{\lambda}$.
Consequently the eigenvalues are
\begin{equation}
E^{\pm}_{0}=\frac{1}{2}\left[\varepsilon_{0}^{0}+\varepsilon_{1}^{-1}\pm\sqrt{\left(\varepsilon_{0}^{0}-\varepsilon_{1}^{-1}\right)^{2}+4\left(R_{0,1}\right)^{2}}\right],
\end{equation}
and the corresponding normalized eigenstates are
\begin{equation}\label{eq:wf_2}
\begin{split}
\left|\tilde{\psi}_{0}^{\pm}\right\rangle   = & \left(\begin{array}{c}
\pm\sqrt{\frac{1}{2}\left(1\pm\frac{\varepsilon_{0}^{0}-\varepsilon_{1}^{-}}{\sqrt{\left(\varepsilon_{0}^{0}-\varepsilon_{1}^{-}\right)^{2}+4\left(R_{0}^{1}\right)^{2}}}\right)}\\
\sqrt{\frac{1}{2}\left(1\mp\frac{\varepsilon_{0}^{0}-\varepsilon_{1}^{-}}{\sqrt{\left(\varepsilon_{0}^{0}-\varepsilon_{1}^{-}\right)^{2}+4\left(R_{0}^{1}\right)^{2}}}\right)}
\end{array}\right).
\end{split}
\end{equation}
For the ground state $\left|\tilde{\phi}_{g}\right\rangle=\left|-1_{x}\right\rangle \otimes \left|0\right\rangle$, the ground-state energy is
\begin{equation}\label{eg1}
E_{g} = \frac{1}{2}\left(\lambda^{2}\omega-2g\lambda\right) -\Omega e^{-\frac{\lambda^{2}}{2}}.
\end{equation}

\begin{figure}[!tb]
	\centering	
	\includegraphics[width=0.9\columnwidth]{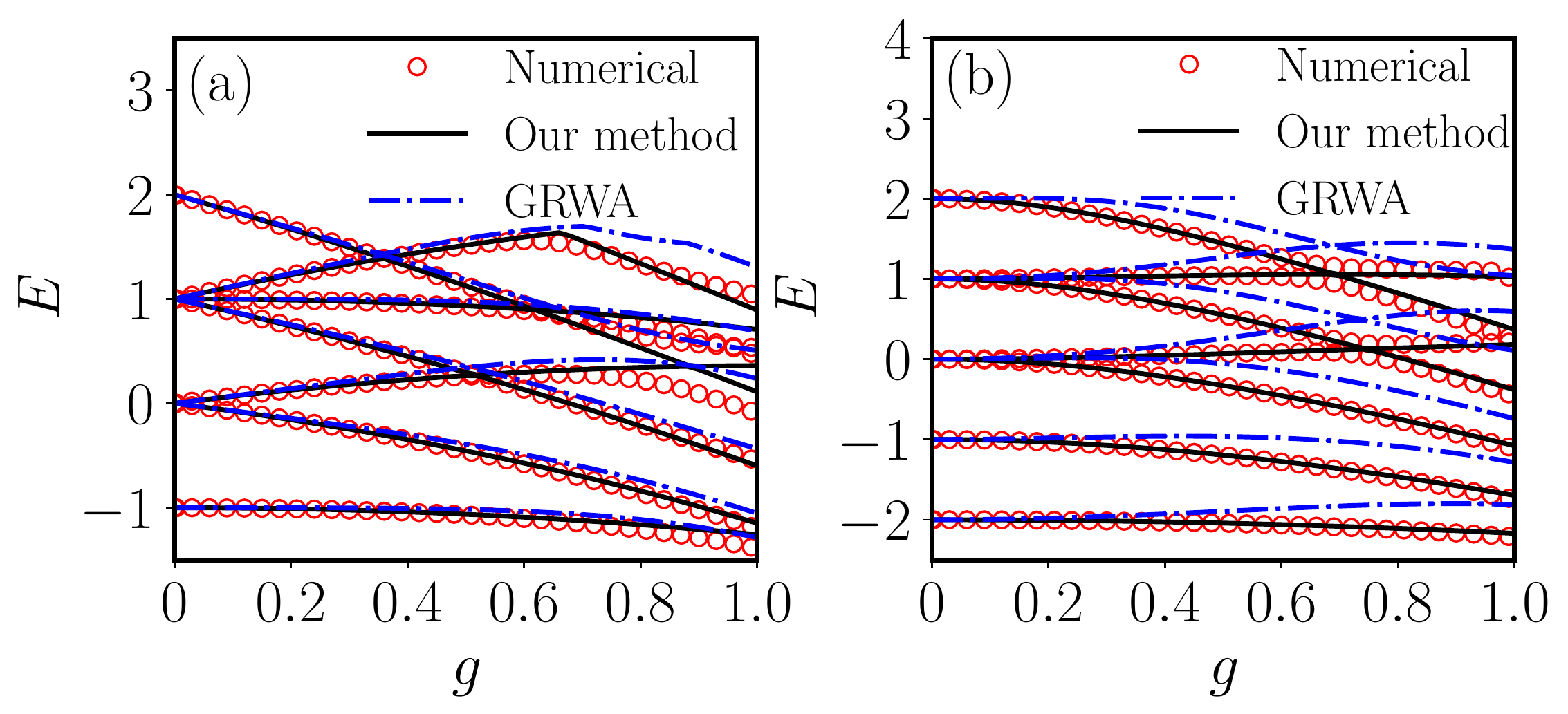}
	\caption{The energy spectra of the system with (a) $\Omega =1.0$ and (b) $\Omega =2.0$. The numerical result is obtained by an exact diagonalization of Hamiltonian Eq. \eqref{eq_hamiltonian}.  The results for our method and the GRWA are obtained by solving Hamiltonian Eq. \eqref{grwah} with an optimal $\lambda = g/(\omega+\Omega)$ and  $\lambda = g/\omega$, respectively.  }\label{fig_energy_spectrum}
\end{figure}

Up to here, the only one task lefted to be completed is determining the parameter $\lambda$. The unitary transformation we employed presents the form of a set of  approximate wave functions. Particularly, we extract the one for the ground state as
\begin{equation}
|\phi_G(\lambda)\rangle = U(\lambda) \left|-1_{x}\right\rangle \otimes \left|0\right\rangle,
\end{equation}
where $U(\lambda)$ is defined as previous one $U(\lambda) = e^{\lambda \hat{J}_z(\hat{a}^{\dagger}-\hat{a})}$. The $|\phi_G(\lambda)\rangle$ can be considered as a trial ground state wave function with undetermined parameter $\lambda$. Thus, the ground state energy function can be readily obtained:
\begin{equation}\label{eg2}
E_G(\lambda)=\langle \phi_G(\lambda)|\hat{H}|\phi_G(\lambda)\rangle= \frac{1}{2}\left(\lambda^{2}\omega-2g\lambda\right) -\Omega e^{-\frac{\lambda^{2}}{2}},
\end{equation}
where $\hat{H}$ is the whole model Hamiltonian. The parameter $\lambda$ can be determined by minimizing the ground state wave function in Eq. \eqref{eg2}, namely $ \partial E_G/\partial \lambda  = 0$ which yields
\begin{equation}\label{eq:GRWA_lambda}
g-\lambda\omega-\lambda\Omega e^{-\lambda^{2}/2}  =  0.
\end{equation}

The approximate solution is then obtained:
\begin{equation}\label{eq:lambda}
\lambda  =   \frac{g}{\omega+\Omega e^{-\lambda_{0}^{2}/2}},
\end{equation}
where $\lambda_{0}=g/\left(\omega+\Omega\right)$. In the small $g$ limit, we can further simplify the solution to

\begin{equation}\label{lambda0}
\lambda = \frac{g}{\omega+\Omega},
\end{equation}
which is the same as $\lambda_{0}$.
Note that a more accurate $\lambda$ value can be acquired by numerically solving Eq.~\eqref{eq:GRWA_lambda}.
In this work, we focus on the analytical study for an intuitive understanding. Substituting the value of $\lambda$ in Eq. \eqref{lambda0} to Eq.(14-19), our method completes.

Although $\lambda$ is obtained in a variational manner for the ground state, we find that our method also improves the conventional GRWA for the excited states. The reason is addressed as follows. Note that the coefficient of the general counter rotating term is
\begin{equation}\label{co_cgrwa}
	g-\lambda\omega-\Omega e^{-\lambda^{2}/2}\frac{\lambda}{n+1}L_{n}^{1}\left(\lambda^{2}\right).
\end{equation}
When coupling strength $\lambda$ is small enough,
using $L_{n}^{1}\left(x\right) \to 1$ if $x\ll1$, Eq.~\eqref{co_cgrwa} can be simplified as
\begin{equation}
g-\lambda\omega-\lambda\Omega e^{-\lambda^{2}/2} ,
\end{equation}
which vanishes when Eq. \eqref{eq:GRWA_lambda} is adopted. It ensures that the approximate Hamiltonian \eqref{grwah} is exact up to single-excitation level since the general counter rotating term vanishes.

\section{The advantage of the variational method}\label{sec_4}
Our variational method improves the GRWA in both quantitative and qualitative ways. We take energy spectra, the mean photon number and the dynamic process as examples to compare the variational method with the GRWA. The numerically exactly diagonalized results are also involved in the comparison as a benchmark.

From the comparison of the energy spectra between different methods, the quantitative advantage of the variational method over the GRWA is clearly revealed. Figure~\ref{fig_energy_spectrum} displays the energy spectra with two sets of parameters, i. e., $\Omega=1.0$ and $\Omega=2.0$ (we set $\omega=1$ as an energy unit). From both panels, we can see that energy spectra calculated by both our method and the GRWA are undistinguishable and agree with the numerical ones well when the coupling strength $g$ is small.
However, as $g$ rises, the difference between the two methods becomes noticeable. It is because the difference between the corresponding variational parameters $\lambda=g/\omega$ and $\lambda=g/(\omega+\Omega)$ increases as $g$ becomes larger. Although the GRWA gains the mean feature of the cross structure of the energy spectra, it evidently deviates from the numerical solution for large $g$. However, our variational method keeps pace with the exact one. We also find that when $\Omega$ becomes large, the GRWA gets worse [see Figs.~\ref{fig_energy_spectrum} (a) and (b)]. It can be readily understood by that the deviation from optimal parameter $\lambda=g/(\omega+\Omega)$ to the GRWA one $\lambda=g/\omega$ exaggerates when $\Omega$ gets larger.

\begin{figure}[!tb]
	\centering
	\includegraphics[width=0.9\columnwidth]{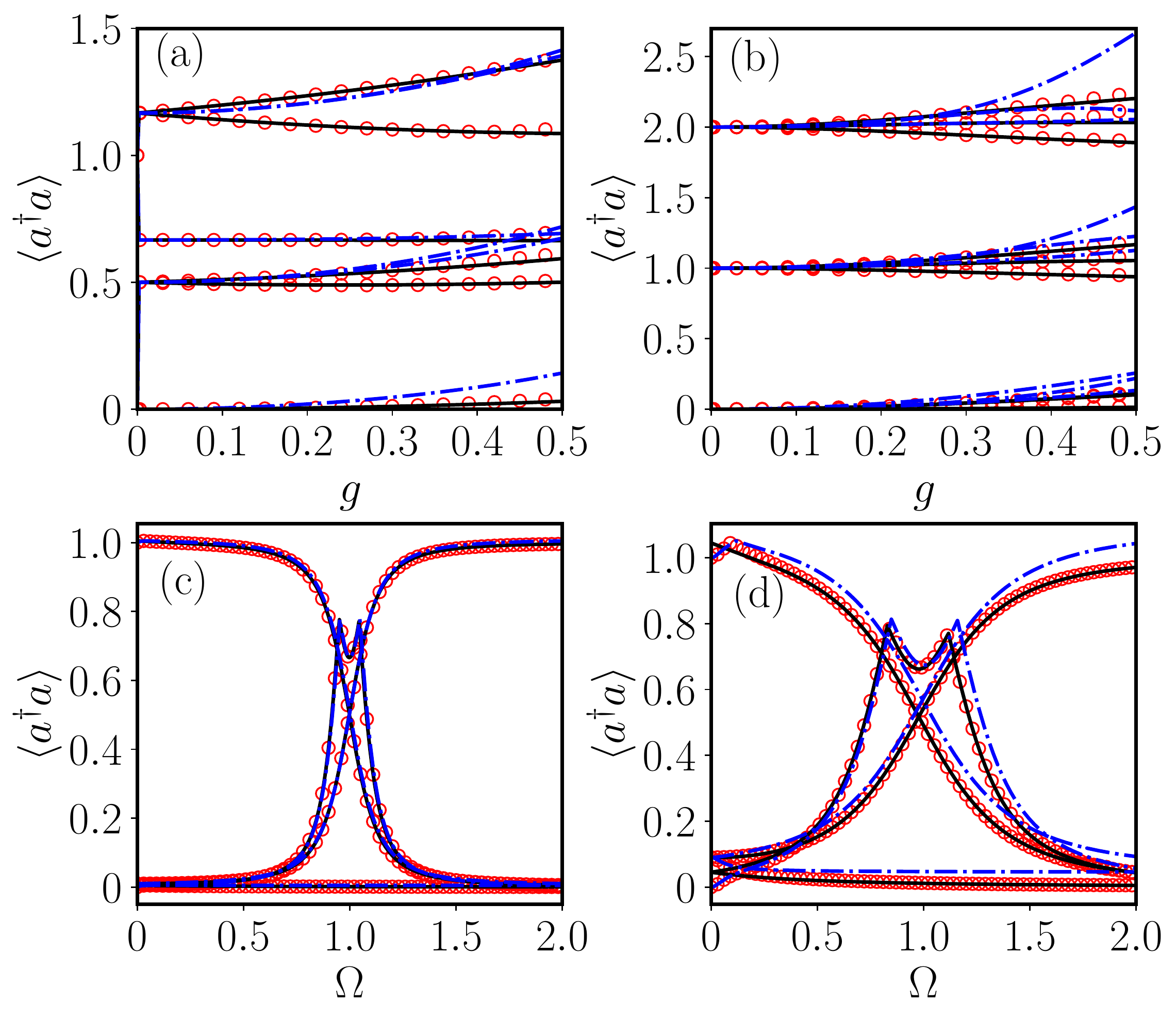}
	\caption{Mean photon number as a function of $g$ for (a) $\Omega=1 $, (b) $\Omega=2$, and as a function of $\Omega$ for (c) $g=0.1$, (d) $g=0.3$. We compare our results (solid line, obtained by Eq. (\ref{app_n0},\ref{app_n12},\ref{app_n3})) with those obtained by the numerical exact diagonalization method (circle) and GRWA (dashed line).}\label{fig_photon}
\end{figure}

Besides the energy spectra, the mean photon number is another fundamental observable in the light-matter interacting systems.
The results indicate the accuracy of our method in calculating the mean photon number in the ground state as well as low excited states.  Figures~\ref{fig_photon} (a) and (b) show the mean photon number $\langle \hat{a}^{\dagger}\hat{a}\rangle$ as a function of the coupling strength $g$. Similar to the energy spectra, we see that our method evidently improves the accuracy of the GRWA, and agrees well with the numerical exact one.
We also show $\langle \hat{a}^{\dagger}\hat{a}\rangle$ as a function of $\Omega$.
The importance of our method becomes more evident as $\Omega$ increases, because the obtained value by the GRWA will have more derivation from the exact one as $\Omega$ enlarges, as can be clearly seen in Figs.~\ref{fig_photon} (c) and (d).

\begin{figure}[!tb]
	\centering
	\includegraphics[width=0.9\columnwidth]{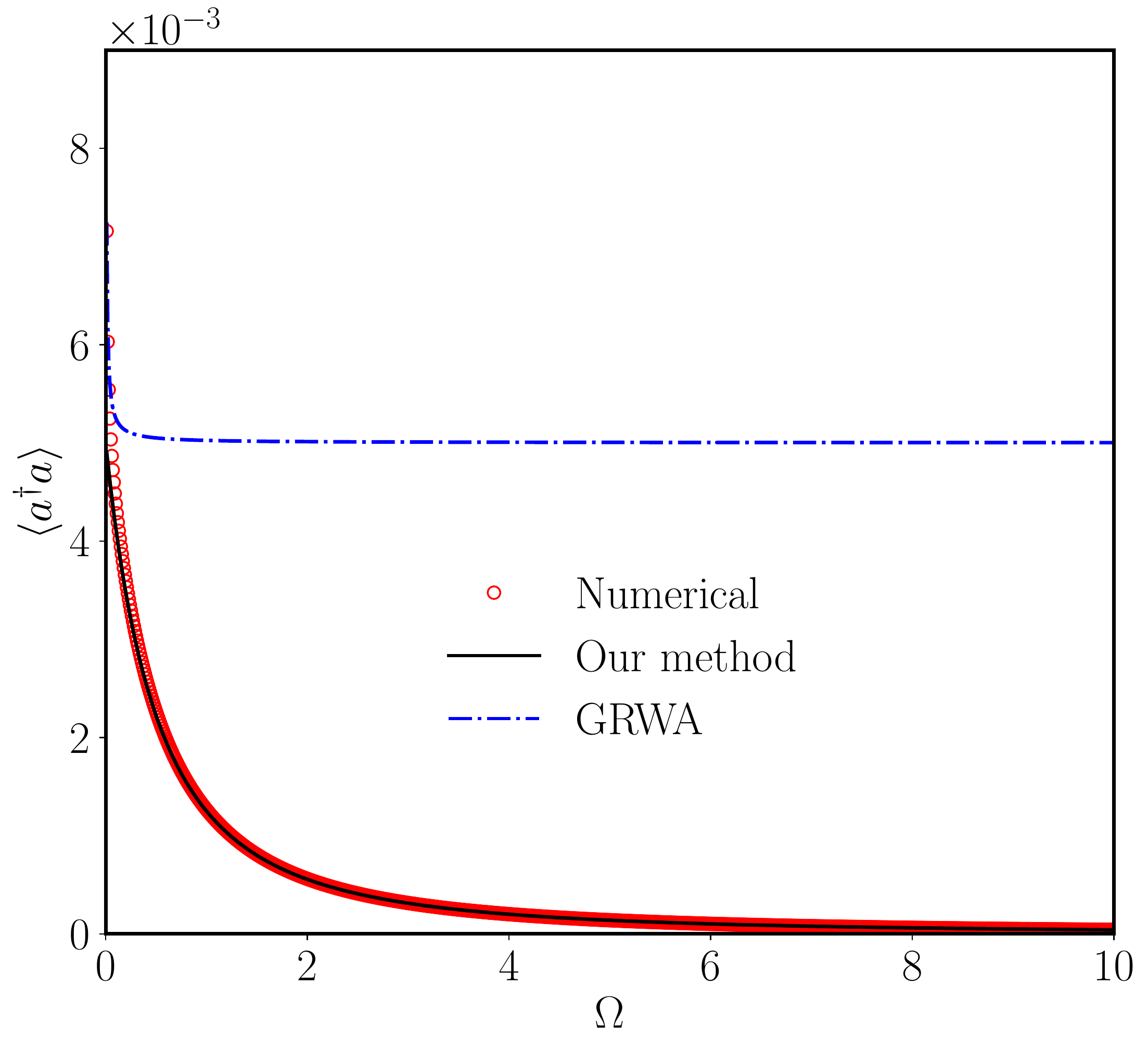}
	\caption{The comparison of mean photon number for ground state obtained by different methods. Here we choose $g=0.1$. }\label{fig_photon_zoomin}
\end{figure}

In the two-qubit model, recall that the mean photon number of the ground state obtained by GRWA~\cite{Zhang2015} is
\begin{equation}\label{eq:photon_grwa}
\left\langle a^{\dagger}a\right\rangle=\frac{1}{2}\left(1+\frac{\chi_{0}}{\sqrt{\chi_{0}^{2}+8}}\right)\frac{g^{2}}{\omega^{2}},
\end{equation}
where $\chi_{0}=\frac{\sqrt{2}g^{2}}{\Omega\omega}e^{\frac{g^{2}}{2\omega^{2}}}$. In our method, the mean photon number of the ground state (see details in Appendix \ref{appendix_photon}) is
\begin{equation}\label{eq:photon_our}
\left\langle\hat{a}^{\dagger}\hat{a}\right\rangle = \frac{\lambda^2}{2},
\end{equation}
where $\lambda={g}/{\left(\omega+\Omega\right)}$. As is shown in Fig. \ref{fig_photon_zoomin}, we find that the mean photon number obtained by GRWA is always larger than $g^{2}/(2\omega^{2})$ while our result is always smaller than $g^{2}/(2\omega^{2})$.

Recently, a novel quantum phase transition in large frequency ratio ($\eta=\Omega/\omega \to \infty$) has been extensively discussed~\cite{Wang2018_053061,Hwang2015,Liu2017}. Note that the large $\eta$ limit is equivalent to the large $\Omega$ limit here since $\omega = 1$ has been taken. In such a condition, the ground state can be categorized into two phases through a phase transition point: one is the normal phase for small coupling strength $g$, where the mean photon number is zero; the other is the superradiant phase for sufficiently large $g$, where the mean photon number is finite. The analytical formulas of Eqs.~\eqref{eq:photon_grwa} and~\eqref{eq:photon_our} allow us to explore the asymptotic behavior in the large $\Omega$ limit. In this limit, if $g$ is small, the value of $\left\langle \hat{a}^{\dagger}\hat{a} \right\rangle$ obtained by our method approaches to zero, while the mean photon number obtained by the GRWA method approaches to $g^{2}/\left(2\omega^{2}\right)$, as is exhibited in Fig.~\ref{fig_photon_zoomin}. The validity of mean photon number in large frequency ratio shows the variational method captures the more essential physics, that is missed by the GRWA.

Apart from the static properties, the dynamical evolution is another significant issue in Rabi physics. In this work,  we study the time evolution of the polarization $\left\langle \hat{J}_{z}\right\rangle$ and the population of the qubits remaining in the initial state $\left|-1_{z}\right\rangle$ as two examples. The two physical quantities can be defined as
\begin{equation}
J_{z}\left(t\right)  =  \left\langle \tilde{\varphi}\left(t\right)\right|\hat{\tilde{J}}_{z}\left|\tilde{\varphi}\left(t\right)\right\rangle,
\end{equation}
and
\begin{equation}
P_{-1}(t)  =  \left\langle -1_{z}\right|\left(\mathrm{Tr_{ph}}\left|\tilde{\varphi}(t)\right\rangle\left\langle\tilde{\varphi}(t)\right|\right)\left|-1_{z}\right\rangle,
\end{equation}
respectively.
Based on our variational method, the dynamic process can be analytically expressed.
The initial state in the original Hamiltonian is chosen as $\left|\varphi\left(0\right)\right\rangle   =  e^{\alpha\left(\hat{a}^{\dagger}-\hat{a}\right)}\left|-1_{z},0\right\rangle$. Our analytical calculation is performed in the transformed frame. Thus, the initial state can be obtained using the unitary transformation $\hat{U}$ as $\left|\tilde{\varphi}(0)\right\rangle=\hat{U}\left|\varphi(0)\right\rangle$. The wavefunction evolves as
{$\left|\tilde{\varphi}\left(t\right)\right\rangle   = e^{-{\rm i}\hat{\tilde{H}}_{\mathrm{GRWA}}t}\left|\tilde{\varphi}\left(0\right)\right\rangle$.}
The detailed formulas of the dynamic process is exhibited in Appendix \ref{appendix_dynamics}.

\begin{figure}[!tb]
	\centering
	\includegraphics[width=\columnwidth]{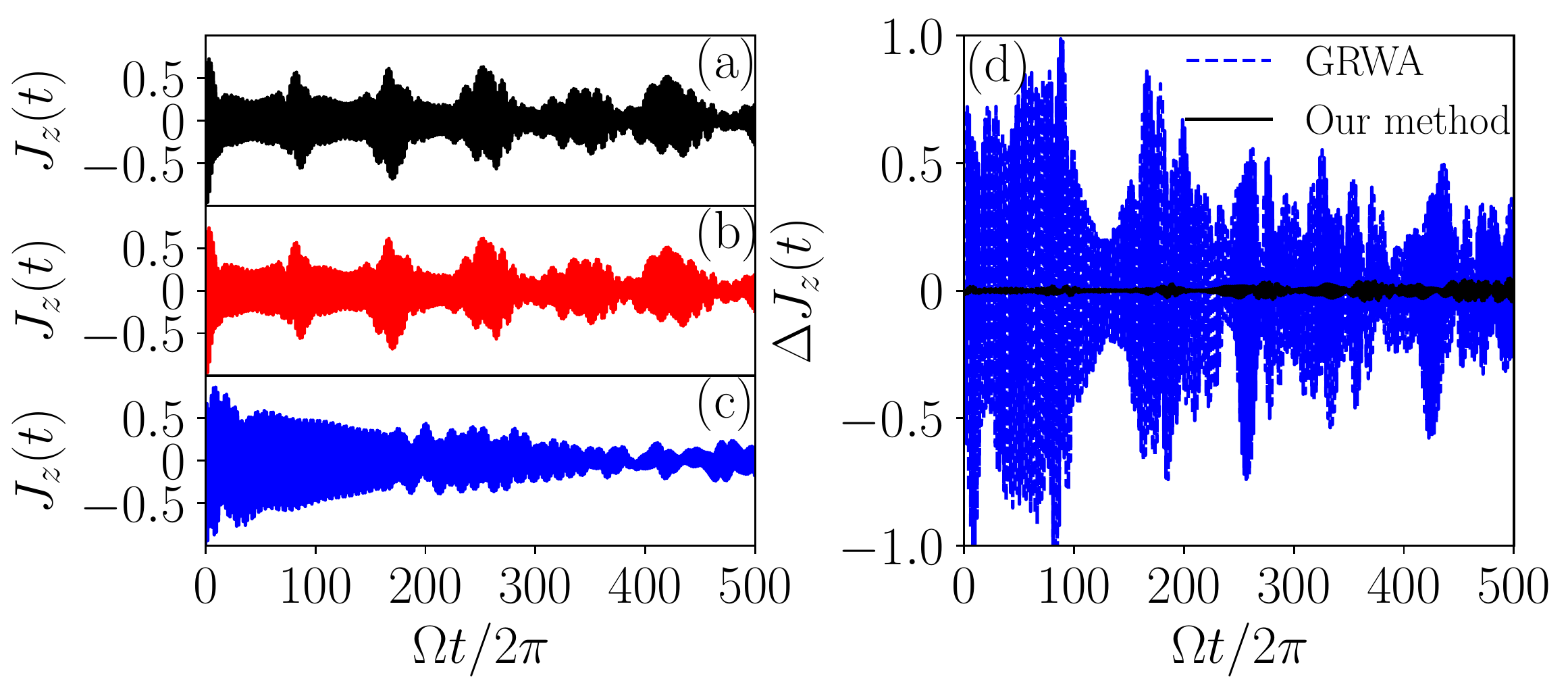}
	\caption{Time evolution of $\left\langle\hat{{J}}_{z}\right\rangle$ with $g=0.2$, $\Omega =2$. For the initial state, we choose $\alpha=2$. (a) The results obtained by our method (see Eq. \eqref{app_jz}). (b) The exact diagonalization results. (c) The GRWA results. (d) The deviations of the analytical results (our method and the GRWA) to the numerical exact one.  }\label{fig_dynamics_W}
\end{figure}

\begin{figure}[!tb]
	\centering
	\includegraphics[width=\columnwidth]{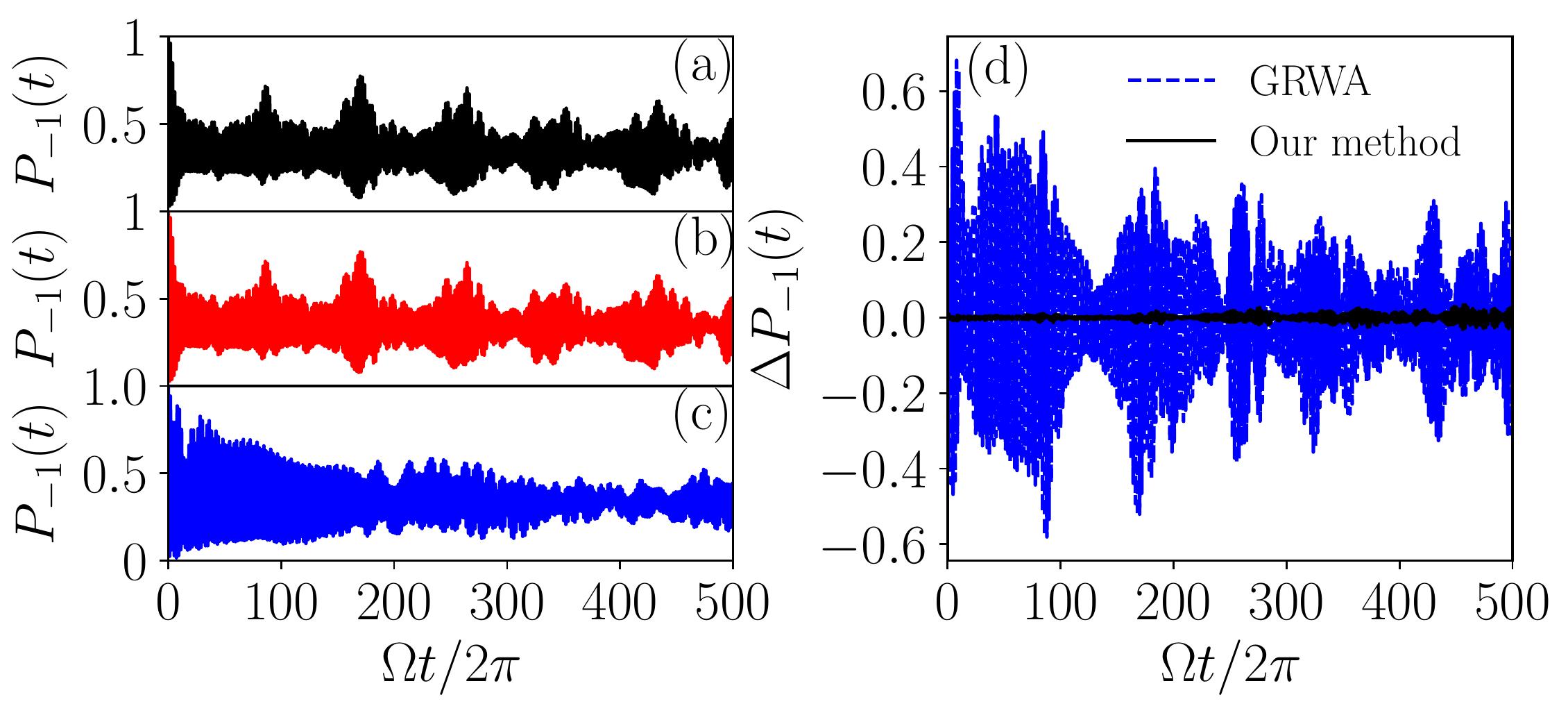}
	\caption{Population of the qubits remaining in the initial state $\left|-1_{z}\right\rangle$ with $g=0.2$, $\Omega =2$. For the initial state, we choose $\alpha=2$. (a) The results obtained by our method (see Eq. \eqref{app_p}). (b) The exact diagonalization results. (c) The GRWA results. (d) The deviations of the analytical results (our method and the GRWA) to the numerical exact one.}\label{fig_dynamics_P}
\end{figure}

Figures~\ref{fig_dynamics_W} and ~\ref{fig_dynamics_P} show $J_{z}(t)$ and $P_{-1}(t)$ respectively. In order to illustrate the improvement of our variational method over the GRWA, the numerical exact result is incorporated as a benchmark.
In order to study the characteristics of the dynamics process, we give the results with evolution time up to $\Omega t/(2\pi)=500$.
Figures~\ref{fig_dynamics_W}(a, b) and ~\ref{fig_dynamics_P}(a, b) show that both $J_{z}(t)$ and $P_{-1}(t)$ obtained by the numerical exact method exhibit obvious quasi-periodical structure.
Comparing to the numerical exact method, the results obtained by our method seize the characteristic panorama, which is missed by the GRWA, as be shown in Figs.~\ref{fig_dynamics_W}(c) and ~\ref{fig_dynamics_P}(c).
Figures~\ref{fig_dynamics_W}(d) and ~\ref{fig_dynamics_P}(d) display the deviations of the analytical results (our method and the GRWA) to the numerical exact one.
From Figs.~\ref{fig_dynamics_W}(d) and~\ref{fig_dynamics_P}(d), we can find that the dynamics processes calculated by our method agrees with the numerical exact ones with high accuracy. In contrast, the GRWA results evidently deviate.
This indicates that our method has obvious improvement over the GRWA.

Based on the results of time evolution, it is obvious that our variational method has an important qualitative correction on the GRWA. Furthermore, considering the fact that dynamical process relates with the energy spectra of the system, the results shown in Figs.~\ref{fig_dynamics_W} and~\ref{fig_dynamics_P} indicate that our method can obtain the energy spectra close to the exact one  with high accuracy.

\section{discussion and conclusions}\label{sec_5}
We have developed an analytical approximation for the two-qubit quantum Rabi model. Although still employing the GRWA frame, we further extend it by introducing the variational method. The advantage of our method is clearly revealed in both quantitative and qualitative aspects. We have explored the energy spectra, the mean photon number and the dynamical processes. The outcome obtained by our method shows good agreement with the exact numerical calculation and evidently improves the accuracy of the GRWA. Furthermore, the importance of the variational method is showed by the correct prediction of the zero mean photon in normal phase and the quasi-periodical structure of time evolution. We also expect our method will be helpful to understand the physics in two-qubit systems, which are quite fundamental to perform the quantum state manipulation, such as quantum state preparation and quantum computing \cite{Romero2012_120501,Wang2016_012328}.

Finally, we would like to discuss the applicable parameter ranges of our approximate method. The main idea of the GRWA can be roughly regarded as to employ the displaced oscillator as the nonperturbed term 
and consider the atomic term $\Omega \hat{J}_x$ as a perturbation. However, we realize that problems readily
come when atom energy scale enlarges and can no longer be taken as a perturbation. 
Luckily, for a large value of $\Omega$ accompanying with a small coupling strength $g$,
the displaced oscillator can still be dominant if the the displacement is renormalized via a variational way, and thus our variational method works in such a case. 
Summarizing, we can efficiently improve the GRWA for arbitrary $\Omega$ with a perturbative $g$. 
Beyond the above mentioned parameter regimes, in principle our method fails. For example, for an intermediate coupling strength $g$ and a large $\Omega$, the ground-state energy obtained by the variational method has a clear deviation from the numerical result.

\section{acknowledgement}
This work is supported by Ministry of Science and Technology of China (Grant No. 2016YFA0301300), National Science Foundation of China (Grant Nos. 11604009, 11474211, 11674139, 11834005) and China Postdoctoral Science Foundation (Grant Nos. 2015M580965 and 2016T90028). We also acknowledge
the computational support in Beijing Computational Science Research Center (CSRC).

%

\appendix
\begin{widetext}

\section{Adiabatic approximation results}\label{appendix_aa}
The matrix shown in Eq.~(7) in the maintext can be easily diagonalized, the results are shown in the following.

The eigenvalues are
\begin{equation}\label{eq:eig_0}
	\tilde{\epsilon}_{n}^{0}\left(\lambda\right)  =  \xi_{n}^{0},
\end{equation}
\begin{equation}\label{eq:eig_pm}
	\tilde{\epsilon}_{n}^{\pm}\left(\lambda\right)
	 =  \frac{1}{2} \left(\xi_{n}^{-}+\xi_{n}^{+}\pm\sqrt{\left(\xi_{n}^{-}-\xi_{n}^{+}\right)^2+4 \left(\epsilon_{\lambda}\right)^2}\right),
\end{equation}
and the corresponding eigenstates are
\begin{equation}\label{eq:AA_matrix}
\left|\tilde{\epsilon}_{n}^{0}\left(\lambda\right)\right\rangle =\left(\begin{array}{c}
0\\
1\\
0
\end{array}\right),
\end{equation}

\begin{equation}
\begin{split}
\left|\epsilon_{n}^{\pm}(\lambda)\right\rangle =\left(\begin{array}{c}
\pm\sqrt{\frac{1}{2}\left[\left(1\pm\frac{\xi_{n}^{-}-\xi_{n}^{+}}{\sqrt{\left(\xi_{n}^{-}+\xi_{n}^{+}\right)^{2}+4\left(\epsilon_{\lambda}\right)^{2}}}\right)\right]}\\
0\\
\sqrt{\frac{1}{2}\left[\left(1\mp\frac{\xi_{n}^{-}-\xi_{n}^{+}}{\sqrt{\left(\xi_{n}^{-}+\xi_{n}^{+}\right)^{2}+4\left(\epsilon_{\lambda}\right)^{2}}}\right)\right]}
\end{array}\right).
\end{split}
\end{equation}
If we want to determine the optimal value of $\lambda$, we should solve $\frac{\partial{\tilde{\epsilon}_{0}^{-}}}{\partial{\lambda}}=0$. An analytical expression can not be readily
obtained as a consequence of cumbersome form.

\section{Energy spectrum of the model calculated by our method}\label{appendix_energy}
For the energy spectra, we can easily obtain three lowest energies. For the Hamiltonian matrix in the $n$-th manifold subspace, we can diagonalize it through the method used by Zhang \textit{et al.}~\cite{Zhang2015}
For simplicity, we write the Hamiltonian matrix in Eq.~\eqref{eq:H_matrix} as
\begin{equation}
\hat{\tilde{H}}^{\prime}_{n}(\lambda)=\left(\begin{array}{ccc}
\nu_{-} & z & 0\\
z & \nu_{0} & y\\
0 & y & \nu_{+}
\end{array}\right),
\end{equation}
where
$\nu_{-}=\omega\left(n-1\right)+f_{n-1}^{0}+\epsilon_{\lambda}$, $\nu_{0}=\omega n+f_{n}^{0}+2\epsilon_{\lambda}$, $\nu_{+}=\omega\left(n+1\right)+f_{n+1}^{0}+\epsilon_{\lambda}$, $z=\sqrt{\frac{n}{2}}\left(f_{n-1}^{1}+\lambda^{\prime}\right)$, $y=\sqrt{\frac{n+1}{2}}\left(f_{n}^{1}+\lambda^{\prime}\right)$.

The determinant is
\begin{equation}
\left|\begin{array}{ccc}
\nu_{-} & z & 0\\
z & \nu_{0} & y\\
0 & y & \nu_{+}-E
\end{array}\right|=0,
\end{equation}
and it gives the cubic equation $E^{3}+bE^{2}+cE+d=0$, where
\begin{equation}
\begin{split}
b=&-\nu_{-}-\nu_{0}-\nu_{+},\\
c=&\nu_{-}\nu_{0}+\nu_{+}\left(\nu_{-}+\nu_{0}\right)-z^{2}-y^{2},\\
d=&-\nu_{-}\nu_{0}\nu_{+}+z^{2}\nu_{+}+y^{2}\nu_{-}.
\end{split}
\end{equation}
Then we can easily obtain three eigenvalues for each $n>0$ as
\begin{equation}
\begin{split}
E_{n}^{1}=&\frac{-b-2\sqrt{b^{2}-3c}\mathrm{cos}\theta}{3},\\
E_{n}^{2}=&\frac{-b+\sqrt{b^{2}-3c}\left(\mathrm{cos}\theta+\sqrt{3}\mathrm{sin}\theta\right)}{3},\\
E_{n}^{3}=&\frac{-b+\sqrt{b^{2}-3c}\left(\mathrm{cos}\theta-\sqrt{3}\mathrm{sin}\theta\right)}{3},
\end{split}
\end{equation}
where $\theta=\frac{1}{3}arc\mathrm{cos}\left[\frac{2b\left(b^{2}-3c\right)-3a\left(bc-9d\right)}{2\sqrt{\left(b^{2}-3c\right)^{3}}}\right]$ when $\left(bc-9d\right)^2-4\left(b^{2}-3c\right)\left(c^{2}-3bd\right)<0$.

The eigenstates are
\begin{equation}
\left|\tilde{\phi}_{n}^{j}\right\rangle=c_{1,n}^{j}\left|1_{x},n-1\right\rangle+c_{0,n}^{j}\left|0_{x},n\right\rangle+c_{-1,n}^{j}\left|-1_{x},n+1\right\rangle,
\end{equation}
where the coefficients are
\begin{equation}
\begin{split}
c_{-1,n}^{j}=&\frac{y\left(E_{n}^{j}-\nu_{-}\right)}{\eta},\\
c_{0,n}^{j}=&\frac{\left(E_{n}^{j}-\nu_{+}\right)\left(E_{n}^{j}-\nu_{-}\right)}{\eta},\\
c_{1,n}^{j}=&\frac{z\left(E_{n}^{j}-\nu_{+}\right)}{\eta},
\end{split}
\end{equation}
with the normalized parameter $\eta^{2}=y^{2}\left(E_{n}^{j}-\nu_{-}\right)^{2}+\left(E_{n}^{j}-\nu_{+}\right)^{2}\left(E_{n}^{j}-\nu_{-}\right)^{2}+z^{2}\left(E_{n}^{j}-\nu_{+}\right)^{2}$.

\section{Mean photon number calculated by our method}\label{appendix_photon}
For the photon number $\left\langle \hat{a}^{\dagger}\hat{a}\right\rangle$, we can obtain it's expression in the transformed representation as
\begin{equation}
\begin{split}
\hat{\tilde{n}} = &\hat{U}\hat{a}^{\dagger}\hat{a}\hat{U}^{\dagger}\\
= & \hat{a}^{\dagger}\hat{a}-\lambda \hat{J}_{z}\left(\hat{a}^{\dagger}+\hat{a}\right)+\lambda^{2}\hat{J}_{z}^{2}\\
= & \hat{a}^{\dagger}\hat{a}-\lambda \frac{\hat{J}_{+}+\hat{J}_{-}}{2}\left(\hat{a}^{\dagger}+\hat{a}\right)+\lambda^{2}\frac{\left(\hat{J}_{+}+\hat{J}_{-}\right)^{2}}{4}.
\end{split}
\end{equation}

Then we can calculate the mean photon number by $\left\langle \hat{\tilde{O}}\right\rangle=\left\langle \tilde{\phi}\right|\hat{\tilde{O}}\left|\tilde{\phi}\right\rangle$. Since the energy spectra and corresponding wavefunctions are calculated in different subspaces, the mean photon number is also calculated in different subspaces.

For the ground state,
\begin{equation}\label{app_n0}
\begin{split}
\langle \hat{\tilde{n}}_{g}\rangle=&\left\langle \tilde{\phi}_{g}\right|\hat{\tilde{n}}\left|\tilde{\phi}_{g}\right\rangle=\frac{\lambda^{2}}{2}.
\end{split}
\end{equation}

For the 2-nd and 3-th excited states
\begin{equation}\label{app_n12}
\begin{split}
\langle \hat{\tilde{n}}^{j}_{0}\rangle=&\left\langle \tilde{\phi}_{0}^{j}\right|\hat{\tilde{n}}\left|\tilde{\phi}_{0}^{j}\right\rangle\\
=&\frac{\lambda^{2}}{2}+\left(\frac{\lambda}{\sqrt{2}}c_{0,0}^{j}-c_{-1,0}^{j}\right)\left[\frac{\lambda}{\sqrt{2}}\left(c_{0,0}^{j}\right)^{*}-\left(c_{-1,0}^{j}\right)^{*}\right].
\end{split}
\end{equation}

For the $n$-th manifold states
\begin{equation}\label{app_n3}
\begin{split}
\langle \hat{\tilde{n}}^{j}_{n}\rangle=&\left\langle \tilde{\phi}_{n}^{j}\right|\hat{\tilde{n}}\left|\tilde{\phi}_{n}^{j}\right\rangle\\
= & \left(n+\frac{\lambda^{2}}{2}\right) +\frac{\lambda^{2}}{2}\left(c_{0,n}^{j}\right)^{*}c_{0,n}^{j}-\left(c_{1,n}^{j}\right)^{*}c_{1,n}^{j}+\left(c_{-1,n}^{j}\right)^{*}c_{-1,n}^{j}\\
& -\frac{\sqrt{n}\lambda}{\sqrt{2}}\left[\left(c_{0,n}^{j}\right)^{*}c_{1,n}^{j}+\left(c_{1,n}^{j}\right)^{*}c_{0,n}^{j}\right] -\frac{\sqrt{n+1}\lambda}{\sqrt{2}}\left[\left(c_{-1,n}^{j}\right)^{*}c_{0,n}^{j}+\left(c_{0,n}^{j}\right)^{*}c_{-1,n}^{j}\right].
\end{split}
\end{equation}

\section{Dynamics calculated by our method}\label{appendix_dynamics}

The initial state in the original representation is set as
\begin{equation}
\begin{split}
\left|\varphi\left(0\right)\right\rangle & =  \left|-1_{z},\alpha\right\rangle\\
& =  e^{\alpha\left(a^{\dagger}-a\right)}\left|0\right\rangle \otimes\left|-1_{z}\right\rangle \\
& =  e^{-\left|\alpha\right|^{2}/2}\sum_{n=0}^{\infty}\frac{\alpha^{n}}{\sqrt{n!}}\left|n\right\rangle \otimes\left|-1_{z}\right\rangle. \\
\end{split}
\end{equation}
The initial state in the transformed representation can be obtained as
\begin{equation}
\begin{split}
\left|\tilde{\varphi}(0)\right\rangle=\hat{U}\left|\varphi(0)\right\rangle
= \left|-1_{z},\alpha-\lambda\right\rangle,
\end{split}
\end{equation}
In the basis of $\left|j_{x}\right\rangle$, the eigenvector of $\hat{J}_{z}$ corresponding to eigenvalue $j_{z}=-1$ is
\begin{equation}
\left|-1_{z}\right\rangle=\frac{1}{2}\left|1_{x}\right\rangle-\frac{1}{\sqrt{2}}\left|0_{x}\right\rangle+\frac{1}{2}\left|-1_{x}\right\rangle.
\end{equation}

So, the initial state in the basis of $\left|j_{x},n\right\rangle$ takes the form
\begin{equation}
\begin{split}
	\left|\tilde{\varphi}\left(0\right)\right\rangle
= & \frac{1}{2}\left|1_{x},\alpha-\lambda\right\rangle-\frac{1}{\sqrt{2}}\left|0_{x},\alpha-\lambda\right\rangle+\frac{1}{2}\left|-1_{x},\alpha-\lambda\right\rangle\\
= & \chi_{0}\left|-1_{x},0\right\rangle +\chi_{0,0}\left|0_{x},0\right\rangle +\chi_{-1,0}\left|-1_{x},1\right\rangle \\
& +\sum_{n=1}^{\infty}\left(\chi_{1,n}\left|1_{x},n-1\right\rangle +\chi_{0,n}\left|0_{x},n\right\rangle +\chi_{-1,n}\left|-1_{x},n+1\right\rangle \right),
\end{split}
\end{equation}
where
$
\chi_{0} =  \frac{1}{2}\zeta_{0}^{\alpha-\lambda}$, $
\chi_{-1,n} =  \frac{1}{2}\zeta_{n+1}^{\alpha-\lambda}$, $
\chi_{0,n} =  -\frac{1}{\sqrt{2}}\zeta_{n}^{\alpha-\lambda}$,
$\chi_{1,n} =  \frac{1}{2}\zeta_{n-1}^{\alpha-\lambda}$,
$\zeta_{n}^{\alpha}  = {e}^{-\left|\alpha\right|^{2}/2}\frac{\alpha^{n}}{\sqrt{n!}}
$.

For simplicity, we define $\left|\tilde{\phi}_{0}^{1}\right\rangle=\left|\tilde{\psi}_{0}^{-}\right\rangle$, $\left|\tilde{\phi}_{0}^{2}\right\rangle=\left|\tilde{\psi}_{0}^{+}\right\rangle$. According to Eq.~\eqref{eq:wf_2}, these wavefunctions take the form
\begin{equation}
\left|\tilde{\phi}_{0}^{j}\right\rangle=c_{0,0}^{j}\left|0_{x},0\right\rangle+c_{-1,0}^{j}\left|-1_{x},1\right\rangle.
\end{equation}

The time evolution of the wave function in the transformed Hamiltonian is
\begin{equation}
\begin{split}
\left|\tilde{\varphi}\left(t\right)\right\rangle   = & {e}^{-{\rm i}\hat{\tilde{H}}_{\mathrm{GRWA}}t}\left|\tilde{\varphi}\left(0\right)\right\rangle \\
= & {e}^{-{\rm i}E_{g}t}D_{0}\left|\tilde{\phi}_{g}\right\rangle +\sum_{j=1}^{2}\left({e}^{-{\rm i}E_{0}^{j}t}D_{0}^{j}\right)\left|\tilde{\phi}_{0}^{j}\right\rangle  +\sum_{j=1}^{3}\sum_{n>0}\left({e}^{-{\rm i}E_{n}^{j}t}D_{n}^{j}\right)\left|\tilde{\phi}_{n}^{j}\right\rangle,
\end{split}
\end{equation}
where $D_{0}=\left\langle \tilde{\phi}_{g}\Big|\tilde{\varphi}\left(0\right)\right\rangle $,
$D_{n}^{j}=\left\langle \tilde{\phi}_{n}^{j}\Big|\tilde{\varphi}\left(0\right)\right\rangle$ and take the form
\begin{equation}
\begin{split}
D_{0} =& \chi_{0},\\
D_{0}^{j} = & c_{0,0}^{j}\chi_{0,0}+c_{-1,0}^{j}\chi_{-1,0},\\
D_{n}^{j} = & c_{1,n}^{j}\chi_{1,n}+c_{0,n}^{j}\chi_{0,n}+c_{-1,n}^{j}\chi_{-1,n}.
\end{split}
\end{equation}

Take an expansion in the basis space $\{\left|j_{x},n\right\rangle\}$, the time evolution of the wave function can be written as
\begin{equation}
\begin{split}
\left|\tilde{\varphi}\left(t\right)\right\rangle
= & \beta_{0}\left|-1_{x},0\right\rangle +\beta_{0,0}\left|0_{x},0\right\rangle +\beta_{-1,0}\left|-1_{x},1\right\rangle \\
& +\sum_{n=1}\left(\beta_{1,n}\left|1_{x},n-1\right\rangle +\beta_{0,n}\left|0_{x},n\right\rangle +\beta_{-1,n}\left|-1_{x},n+1\right\rangle \right),
\end{split}
\end{equation}
where
$\beta_{0}  =  {e}^{-{\rm i}E_{g}t}D_{0}$, $\beta_{0,0}  =  \sum_{j=1}^{2}\left( e^{-{\rm i}E_{0}^{j}t}D_{0}^{j}c_{0,0}^{j}\right)$ and $\beta_{-1,0}  =  \sum_{j=1}^{2}\left({e}^{-{i}E_{0}^{j}t}D_{0}^{j}c_{-1,0}^{j}\right)$.

$J_{z}$ in the transformed Hamiltonian can be obtained as $\hat{\tilde{J}}_{z} = \hat{U}\hat{J}_{z}\hat{U}^{\dagger}=\hat{J}_{z}$. Time evolution of $\left\langle \hat{{J}}_{z}\right\rangle$ can be calculated as
\begin{equation}\label{app_jz}
\begin{split}
W\left(t\right)   = & \left\langle \tilde{\varphi}\left(t\right)\left|\hat{\tilde{J}}_{z}\right|\tilde{\varphi}\left(t\right)\right\rangle \\
= & \frac{1}{\sqrt{2}}\left(\beta_{1,1}^{*}\beta_{0,0}+\beta_{0,0}^{*}\beta_{1,1}\right)+\frac{1}{\sqrt{2}}\left(\beta_{0,0}^{*}\beta_{0}+\beta_{0}^{*}\beta_{0,0}\right)\\
& +\sum_{n>0}\left[\frac{1}{\sqrt{2}}\left(\beta_{1,n+1}^{*}\beta_{0,n}+\beta_{0,n}^{*}\beta_{1,n+1}\right) +\frac{1}{\sqrt{2}}\left(\beta_{-1,n-1}^{*}\beta_{0,n}+\beta_{0,n}^{*}\beta_{-1,n-1}\right)\right].
\end{split}
\end{equation}
	
The population for the qubits remaining in the initial state $\left|-1_{z}\right\rangle$ is
\begin{equation}\label{app_p}
\begin{split}
P_{-1}(t)=&\left\langle -1_{z}\right|\left(\mathrm{Tr_{ph}}\left|\tilde{\varphi}(t)\right\rangle\left\langle\tilde{\varphi}(t)\right|\right)\left|-1_{z}\right\rangle\\
=& \frac{1}{4}\beta_{1,1}\beta_{1,1}^{*}+\frac{1}{2}\beta_{0,0}\beta_{0,0}^{*}+\frac{1}{4}\beta_{0}\beta_{0}^{*} +\frac{1}{4}\left(\beta_{0}\beta_{1,1}^{*}+\beta_{1,1}\beta_{0}^{*}\right)\\
& -\frac{1}{2\sqrt{2}}\left(\beta_{0,0}\beta_{1,1}^{*}+\beta_{1,1}\beta_{0,0}^{*}+\beta_{0,0}\beta_{0}^{*}+\beta_{0}\beta_{0,0}^{*}\right)\\
& +\sum_{n>0}\left(\frac{1}{4}\beta_{1,n+1}\beta_{1,n+1}^{*}+\frac{1}{2}\beta_{0,n}\beta_{0,n}^{*}+\frac{1}{4}\beta_{-1,n-1}\beta_{-1,n-1}^{*}\right)\\
& -\sum_{n>0}\frac{1}{2\sqrt{2}}\left(\beta_{0,n}\beta_{1,n+1}^{*}+\beta_{1,n+1}\beta_{0,n}^{*}+\beta_{0,n}\beta_{-1,n-1}^{*}+\beta_{-1,n-1}\beta_{0,n}^{*}\right)\\
& +\sum_{n>0}\frac{1}{4}\left(\beta_{-1,n-1}\beta_{1,n+1}^{*}+\beta_{1,n+1}\beta_{-1,n-1}^{*}\right).
\end{split}
\end{equation}

\end{widetext}

\end{document}